# Constraining Co-Varying Coupling Constants from Globular Cluster Age

**Rajendra P. Gupta**

Department of Physics, University of Ottawa, Ottawa, ON K1N 6N5, Canada;
rgupta4@uottawa.ca

**Abstract:** Equations governing the evolution of a star involve multiple coupling constants. Thus, the time it spends as a main-sequence star can be expected to depend on whether or not such constants vary over the time scale of stellar evolution. When the star belongs to a globular cluster, the star's age cannot exceed that of the globular cluster, and the latter cannot exceed the age of the Universe. This fact can be used to constrain or verify the variation of the coupling constants, i.e., the speed of light $c$, the gravitational constant $G$, the Planck constant $h$, and the Boltzmann constant $k$. We have estimated the age of the main-sequence star analytically from the time it takes to synthesize all its hydrogen into helium under fixed and varying coupling constants scenarios. When we permitted the interrelated variation of the four constants ($G \sim c^3 \sim h^3 \sim k^{3/2}$) and differentiated between the *cosmological* energy and *local* energy conservation laws, we could show that the variation of the constants established in our earlier studies, i.e., $\dot{G}/G = 3\dot{c}/c = 3\dot{h}/h = 1.5\dot{k}/k = 3.90(\pm 0.04) \times 10^{-10}$ yr$^{-1}$ at the current cosmic time is consistent with the present work. Nevertheless, the challenge remains to come up with an experiment, astrometric or terrestrial, that can unequivocally prove or falsify the predicted variation.

## 1. Introduction

The most studied fundamental constants for their potential variations are the dimensionless fine structure constant $\alpha$ and the dimensionful gravitational constant $G$, especially since Dirac [1] predicted their variations by analyzing his large number hypothesis. They have been studied copiously, theoretically as well as observationally. Uzan [2,3] has thoroughly reviewed variations of these constants, among others, including the speed of light $c$, the Planck constant $h$, and the Boltzmann constant $k$. Our interest in this study is in the variations of $c$, $G$, $h$, and $k$. We will, therefore, briefly state their current status without attempting to be comprehensive.

Based on the work of Jordan [4], Brans and Dicke [5] developed a scalar-tensor theory of gravitation wherein $1/G$ was raised to the status of a scalar field potential that could vary spatially and temporally. It was Teller [6] who first attempted to determine a constraint on the variation of $G$ from the stellar scaling laws applied to the evolution of Solar luminosity and the environment required for the existence of life on Earth in the past. Many methods have been developed since then to determine the variation of $G$, which have all resulted in its variation well below Dirac's prediction [1]. These include methods based on solar evolution [6–8], lunar occultation and eclipses [9], paleontological evidence [10], white dwarf cooling and pulsation [11–13], galaxy cluster and globular cluster dimensions [14], age of globular clusters [15], neutron star masses and ages [16], cosmic microwave background temperature anisotropies [17,18], big-bang nucleosynthesis abundances [19,20], asteroseismology [21], lunar laser ranging [22,23], the evolution of planetary orbits [24–26], binary pulsars [27–29], supernovae type Ia luminosity evolution [30–32], and gravitational waves [33–35].

Despite Einstein developing his ground-breaking theory of special relativity based on the constancy of the speed of light, he did consider its potential variation [36]. It was followed by the varying speed of light theories by Dicke [37], Petit [38], and Moffatt [39,40]. Albrecht and Magueijo [41] and Barrow [42] developed such a theory in which Lorentz invariance is broken as there is a preferred frame in which scalar field is minimally coupled to gravity. Other proposals include locally invariant theories [43,44] and vector field theories that cause spontaneous violation of Lorentz invariance [45]. Efforts for constraining the variation of the speed of light include (1) Qui et al. [46]—using supernovae type Ia observations, powerlaw $c$ variation, Hubble parameter $H(z)$, BAO (baryonic acoustic oscillations), and CMB (cosmic microwave background); (2) Salzano et al. [47]—using angular diameter distance $D_A(z)$ maximum, $H(z)$, simulated data, and BAO; (3) Cai et al. [48]—using $H(z)$ and luminosity distance $D_L(z)$ independently determined by Suzuki et al. [49]; (4) Cao et al. [50]—using $D_A(z)$ measurement radio quasars at redshift $z = 1.7$; (5) Cao et al. [51]—suggesting the use of gravitational lensing with background sources, such as supernovae type Ia and quasars; (6) Lee [52]—with statistical analysis of gravitational lensing data on velocity dispersion measurements of 161 systems; and (7) Mendonca et al. [53]—using galaxy clusters mass-fraction measurements.

The Planck constant and the Boltzmann constants are two more constants that play essential roles in our research. It is worth mentioning the efforts of Mangano et al. [54], de Gosson [55], and Dannenberg [56] in considering the variation of the Planck constant $h$. The time-dependent stochastic fluctuations of $h$ were studied by Mangano et al. [54]. The effect of varying $h$ on mixed quantum states was considered by de Gosson [55]. Dannenberg [56] reviewed the relevant literature and elevated $h$ to the status of a dynamical field, which through the Lagrangian density derivatives, couples to other fields and to itself. The Doppler broadening of absorption lines in thermal equilibrium can provide a direct measurement of the Boltzmann constant $k$ e.g., [57,58]. The spectral line profile analysis of quasars and interstellar media should be able to constrain the $k$ variation.

Studying the possible variation of one coupling constant, while keeping fixed other coupling constants shown to be correlated through one common dimensionless function [59], may not be prudent. It likely yields erroneous results when several constants are present in the equations used for the analysis of data. Therefore, we permit in our studies the simultaneous variation of all the interrelated constants $c, G, h$, and $k$. As a result, we were able to (i) resolve the primordial lithium problem [60], (ii) find a reasonable solution to the faint young Sun problem [61], (iii) show that orbital timing studies do not constrain the variation of $G$ [62], (iv) prove that gravitational lensing cannot determine the variation of $c$ [63], (v) establish that SNe Ia data are consistent with the co-varying coupling constants (CCC) model [64], (vi) attest the consistency of the CCC model with bright and extreme quasar Hubble diagrams [65] as well as gamma-ray burst data [66], and (vii) show from local energy conservation the interrelationship $G \sim c^3 \sim h^3 \sim k^{3/2}$ among the constant [59]. Cuzinatto et al. [67] have theoretically confirmed $G \sim c^3$. They considered a scalar-tensor theory of gravity wherein the scalar field $\phi$ includes the gravitational coupling $G$ and the speed of light $c$, both of which are allowed to be functions of the spacetime coordinates.

Our work on the resolution of the faint young Sun problem [61] has special relevance to this work as it still remains amicably unresolved using standard stellar evolution models. These models suggest that the solar luminosity several billion years ago was significantly lower than today, and thus the Earth's temperature would be too low for life to evolve as it is today.

It is a standard practice to apply local energy conservation laws to the problems involving evolutions at a cosmological time scale when it is well known that energy is not conserved in general relativity and cosmology [68–74]. Any conclusion reached due to this practice should thus be treated with caution and not considered proof against Dirac's prediction. As in our recent study towards resolving the faint young Sun problem [61], we will deviate from this practice in this work as well, i.e., we will distinguish between *cosmological* energy and *local* energy (discussed in Section 4).

Our focus in this paper is to explore how the variation of physical constants affects the stellar ages in globular clusters. For this, we will closely follow the analytical approach of Degl'Innocenti et al. [15] and Gupta [61] summarized in Section 2. The results from this approach are presented in Section 3. Section 4 comprises discussion, and Section 5 states our conclusion.

## 2. Analytical Background

Degl'Innocenti et al. [15] expressed in the context of the main-sequence (MS) stellar evolution on the H–R diagram: 'If gravity varies in time, the modification of the MS evolutionary time-scale can be estimated by a surprisingly accurate analytical method.' Accordingly, the MS luminosity $L$ of a star of a given mass and initial elemental composition is determined by the gravitational constant $G$ and the age of the star. Such a star brightens as it burns hydrogen to helium in its core. Denoting the helium abundance in the central region of the star where hydrogen burns as $Y$, we could approximately write

$$L \propto y(Y)g(G) \qquad (1)$$

where $y$ and $g$ are some functions of $Y$ and $G$, respectively. Since helium is produced at a rate proportional to $Y$,

$$L \propto \frac{dY}{dt} \propto y(Y)g(G) \Rightarrow \frac{dY}{y(Y)} \propto g(G(t))dt \qquad (2)$$

Thus, for a star that is born at $t = t_{init}$ with helium abundance $Y_{init}$ (= 0.25) and turns off the MS today ($t = t_0$) with $Y \approx 1$ at its center, we may write

$$\int_{Y_{init}}^{1} \frac{dY}{y(Y)} \propto \int_{t_{init}}^{t_0} dt\, g(G(t)) \qquad (3)$$

Degl'innocenti et al. [15] assumed $g$ to be a power-law function given by $g = (G(t)/G_0)^\gamma$ and determined $\gamma = 5.6$. Therefore, they could write the relation between the apparent turnoff age of a star $\tau_*$ corresponding to the constant $G$ scenario, and the turnoff age of the star $\tau$ with time-dependent $G$, as follows:

$$\tau_* = \int_{t_0-\tau}^{t_0} dt \left(\frac{G(t)}{G_0}\right)^\gamma \qquad (4)$$

Assuming a linear dependence of $G$ on time given by $G(t) = G_0(1 + \Gamma_0(t - t_0))$, writing $p = \gamma + 1$ (= 6.6), integrating Equation (4), and rearranging, they could write

$$\frac{\tau}{\tau_*} = \frac{p\Gamma_0 \tau}{1 - (1 - \Gamma_0 \tau)^p} = \frac{1 - (1 - p\Gamma_0 \tau_*)^{1/p}}{\Gamma_0 \tau_*} \qquad (5)$$

They also considered the power-law variation of $G$, i.e., $G(t) = G_0(t/t_0)^\beta$ and obtained

$$\frac{\tau}{\tau_*} = \frac{(1 + \beta\gamma)\tau/t_0}{1 - (1 - \tau/t_0)^{1+\beta\gamma}} \qquad (6)$$

The authors then analyzed Equations (5) and (6) to establish the bounds on $\dot{G}/G$, i.e., on the $G$ variation, that would yield reasonable main-sequence stellar ages.

We will modify the above approach in line with our recent work related to resolving the faint young Sun problem [61]. The luminosity in that work for a star of mass $M$ and radius $R$, following Newman and Rood [75], is expressed as

$$L \sim \frac{\sigma}{\kappa_0} M^{5.5} R^{-0.5} \left(\frac{G\mu m_H}{k}\right)^{7.5} \qquad (7)$$

Here $\sigma = 2\pi^5 k^4/15c^2 h^3$ is the Stefan–Boltzmann constant, $m_H$ is the mass of a hydrogen atom, $\mu$ is the mean particle mass measured as a fraction of $m_H$, and the opacity is defined by Kramer's opacity law $\kappa = \kappa_0 \rho T^{-3.5}$, with $\rho$ being the density and $T$ the temperature.

The stellar energy equation may be written as [76] (p. 23)

$$\frac{d}{dt}\left(E_{kin} + E_g + E_i + E_n\right) + L_b + L_\nu = 0 \tag{8}$$

The net stellar energy loss due to the kinetic energy $E_{kin}$, the gravitational energy $E_g$, the internal energy $E_i$, and the nuclear energy $E_n$, results in a stellar luminosity that comprises the bolometric luminosity $L_b$ and the neutrino luminosity $L_\nu$. All these energies and luminosities are assumed to be *local*. However, in the CCC approach, we must also consider the change in *cosmological* energy $E_c$ and the corresponding luminosity $L_c$ (Section 4 below and [61]).

Taking the logarithmic time derivative of Equation (7) yields:

$$\frac{\dot{L}}{L} = \frac{\dot{\sigma}}{\sigma} - \frac{\dot{\kappa}_0}{\kappa_0} + 5.5\frac{\dot{M}}{M} - 0.5\frac{\dot{R}}{R} + 7.5\frac{\dot{G}}{G} + 7.5\frac{\dot{\mu}}{\mu} - 7.5\frac{\dot{k}}{k}$$
$$\equiv \frac{\dot{L}_b}{L_b} + \frac{\dot{L}_\nu}{L_\nu} + \frac{\dot{L}_c}{L_c}. \tag{9}$$

Here variations of all the physical constants $c$, $G$, $h$, and $k$, which are explicitly or implicitly contained in equation (9) relate to changes in the unobservable cosmological luminosity $L_c$: $\dot{L}_c/L_c = \dot{\sigma}/\sigma - \dot{\kappa}_0/\kappa_0 + 7.5\dot{G}/G - 7.5\dot{k}/k$. For example, changes in the unobservable cosmological luminosity include changes in the rest-mass energy of the Sun due to changes in the speed of light, but it is not reflected in Equation (7) and thus not considered. Since this energy is *cosmological*, its inclusion will not affect our findings.

Let us now examine the other terms in Equation (9), which are relevant to the observable parameters:

$$\frac{\dot{L}_b}{L_b} + \frac{\dot{L}_\nu}{L_\nu} = 5.5\frac{\dot{M}}{M} - 0.5\frac{\dot{R}}{R} + 7.5\frac{\dot{\mu}}{\mu} \tag{10}$$

1. The solar mass-loss rate due to nuclear fusion is $\approx 7 \times 10^{-14}$ yr$^{-1}$, and that due to solar winds is $\approx 2 \times 10^{-14}$ yr$^{-1}$ [77], giving a total mass-loss rate of $\dot{M}/M \approx -9 \times 10^{-14}$ yr$^{-1}$. Similar mass loss is expected from main-sequence stars of interest in this work. It was shown to be negligible.
2. In the CCC approach, we have $G \sim c^3 \sim h^3 \sim k^{3/2}$. Additionally, $R$ is measured in units of $c$ in relativity and in the CCC model, i.e., $R \rightarrow R_s c/c_0$. We may therefore write $\dot{R}/R \rightarrow \dot{R}_s/R_s + \dot{c}/c$. Here, $R_s$ corresponds to $R$ in the standard model. Since $R_s \sim M$ as per the stellar scaling laws [78], we could expect $\dot{R}_s/R_s \approx \approx -9 \times 10^{-14}$, i.e., the change in $R_s$ for main-sequence stars in the standard model can also be considered negligible see also [75]. We may, therefore, write $\dot{R}/R = \dot{c}/c$.
3. We have to now focus on the last term $\dot{\mu}/\mu$. Taking $X$ as the mean mass fraction of hydrogen (assumed to be 0.75 initially when the star became a main-sequence star) and $Z$ as the mean mass fraction of elements heavier than helium, $\mu$ may be written [79] (p. 54); [78] (p. 42):

$$\frac{1}{\mu} = 2X + \frac{3}{4}Y + \frac{1}{2}Z \tag{11}$$

where $Y = 1 - X - Z$ is the mean mass fraction of helium, and electrons are included in determining the mean particle mass. Since $Z \ll X$, we have:

$$\mu \cong (2 - 1.25Y)^{-1} \tag{12}$$

We may now write the luminosity function that includes only the factors in Equation (7) that are relevant in determining the stellar evolution, i.e., $R$ and $\mu$,

$$L \propto R^{-0.5}(2 - 1.25Y)^{-7.5} \propto dY/dt \tag{13}$$

Here the last proportionality is from Equation (2). Upon writing $R \propto c(t)$ from above, we have

$$\int_{0.25}^{1} dY(2 - 1.25Y)^{7.5} \propto \int_{t_{init}}^{t_0} dt/\sqrt{c(t)} \tag{14}$$

The left-hand side of the equation yields a fixed value (= 8.03) and thus is irrelevant for our comparative analysis. Thus, instead of Equation (4), we may write

$$\tau_* = \int_{t_0-\tau}^{t_0} dt \left(\frac{c(t)}{c_0}\right)^{-1/2} \tag{15}$$

If we assume $c(t)$ evolves linearly, i.e., $c(t) = c_0(1 + c_1(t - t_0))$, we obtain

$$\tau_* = \frac{2}{c_1}[1 - (1 - c_1\tau)^{1/2}], \text{ or}$$

$$\frac{\tau}{\tau_*} = \frac{c_1\tau/2}{1 - (1 - c_1\tau)^{1/2}} = \frac{1 - (1 - c_1\tau_*/2)^2}{c_1\tau_*} \tag{16}$$

However, in the CCC theory, the constants do not evolve linearly with time. Their evolution is typically expressed in terms of the scale factor $a$ or redshift $z$. For the speed of light, $c = c_0 \exp(a^{1.8} - 1)$. Following our earlier work e.g., [61,64], when translated into time variation with the time unit of Gyr, it can be graphically depicted as in Figure 1. We could then write

$$\left(\frac{c(t)}{c_0}\right)^{-1/2} = 1 + 0.0651(t_0 - t) - 0.0013(t_0 - t)^2 \tag{17}$$

It follows from Equation (15) that

$$\frac{\tau_*}{\tau} = 1 + 0.0326\tau - 0.00043\tau^2 \tag{18}$$

The reason we have used the Newman and Rood's exponent of 7.5 on $G$ rather than the more recent exponent of 5.6 determined by Degl'innocenti is that the former yields even stronger constraints on the constants' variation than the latter. So, if our constraint is good for the Newman and Rood's exponent, it would be good for Degl'innocenti equations as well. Another reason to consider Newman and Rood's equations is that it explicitly includes other relevant constants also, and their treatment cannot be ignored in an analysis where these constants are also varying.

We are now ready to compare the findings of Degl'Innocenti et al. [15] with ours.

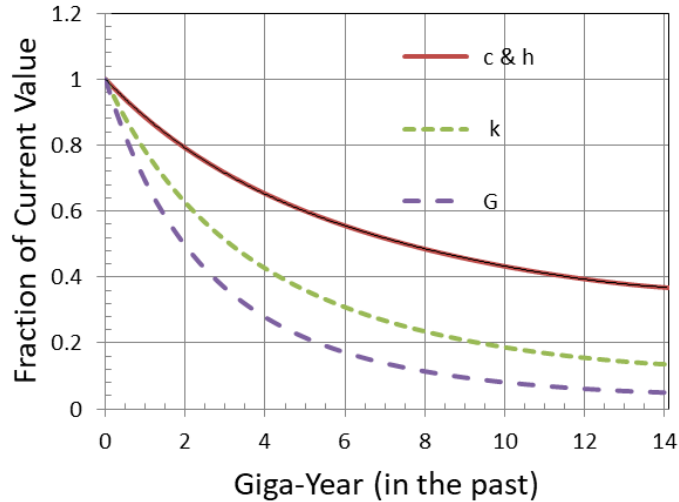

**Figure 1.** Variation of $c, G, h,$ and $k$ with cosmic time. Only $c$ and $h$ vary similarly. $G$ and $k$ vary differently. However, variations of all these constants are interrelated through a common function [59] and not arbitrarily forced.

## 3. Results

In order for the results of the two findings to be comparable, we have to present them in the same format. Thus, even though the CCC results relate to the variation of the speed of light rather than the gravitational constant, we will use the relationship $G \sim c^3$ to convert the $c$ variation to the $G$ variation: $\dot{G}/G = 3\dot{c}/c$.

Let us first consider the calculations of Degl'Innocenti et al. [15]. For the linear dependence of $G$ on time $G(t) = G_0(1 + \Gamma_0(t - t_0))$, we have $\dot{G}_0/G_0 = \Gamma_0$, and for the power-law variation of $G$, i.e., $G(t) = G_0(t/t_0)^\beta$, we obtain $\dot{G}_0/G_0 = \beta/t_0$. They assumed that the star in a GC was born possibly about a Gyr or so after the Big Bang ($t_0 = 14$ Gyr), and therefore they set $/t_0 \approx 1$. We may thus write Equations (5) and (6) as follows:

$$\frac{\tau}{\tau_*} = \frac{\left(1 - \left(1 - 6.6\tau_* \dot{G}_0/G_0\right)^{1/6.6}\right)}{\tau_* \dot{G}_0/G_0} \tag{19}$$

$$\tau/\tau_* = (1 + 5.6 t_0 \dot{G}_0/G_0) \tag{20}$$

By substituting $x \equiv \tau_* \dot{G}_0/G_0$, and $\tau/\tau_* = \mathcal{R}$, Equation (19) may be written as

$$(1 - \mathcal{R}x)^{6.6} = 1 - 6.6x \tag{21}$$

If we consider $0.5 \leq \mathcal{R} \leq 1.5$, i.e., $\tau$ could be 50% lower or 50% higher than $\tau_*$, equation (21) yields real values $x = -0.487$ for $\mathcal{R} = 0.5$ and $x = 0.099$ for $\mathcal{R} = 1.5$. Taking $\tau_* = 14$ Gyr, and since $x \equiv \tau_* \dot{G}_0/G_0$, we determine $-35 \times 10^{-12} \text{yr}^{-1} \leq \dot{G}_0/G_0 \leq 7 \times 10^{-12} \text{yr}^{-1}$ when $G$ variation is linear. Solution of Equation (20) for the same bounds on $\mathcal{R}$, i.e., on $\tau/\tau_*$, and $t_0 = 14$ Gyr, yields $-6 \times 10^{-12} \text{yr}^{-1} \leq \dot{G}_0/G_0 \leq 6 \times 10^{-12} \text{yr}^{-1}$ when $G$ variation follows a power-law.

One might ask, why consider the age ratio $\mathcal{R}$ corresponding to varying $G$ versus constant $G$ as $0.5 \leq \mathcal{R} \leq 1.5$? We could have chosen other limits, but these are more restrictive than those used by Degl'Innocenti et al. in their work [15] and thus would provide more conservative constraints than their work.

Consider now the CCC model. For the linear variation of $c(t) = c_0(1 + c_1(t - t_0)) \Rightarrow \dot{c}_0/c_0 = c_1$, and since $\dot{c}_0/c_0 = \dot{G}_0/3G_0$, Equation (16) becomes

$$\frac{\tau}{\tau_*} = \frac{1 - \left(1 - 0.5\tau_* \dot{G}_0/3G_0\right)^2}{\tau_* \dot{G}_0/3G_0} \tag{22}$$

Following the same steps as above with $x \equiv \tau_* \dot{G}_0/3G_0$, we determine for this case $-429 \times 10^{-12} \text{yr}^{-1} \leq \dot{G}_0/G_0 \leq 429 \times 10^{-12} \text{yr}^{-1}$ for $0.5 \leq \mathcal{R} \leq 1.5$. The constraint for the CCC case is substantially more lax compared to the other case.

We would like to consider now what ratio of $\mathcal{R}$, i.e., $\tau/\tau_*$ we obtain when we use the previously determined value of $\dot{G}_0/G_0 = 390(\pm 4) \times 10^{-12}$ yr$^{-1}$ contained in equation (18) [60,61,64,65]. This equation yields $\tau/\tau_* = 0.73$ to 1, well within the bound considered above. It is also depicted in Figure 2 for different values of the main-sequence stellar ages.

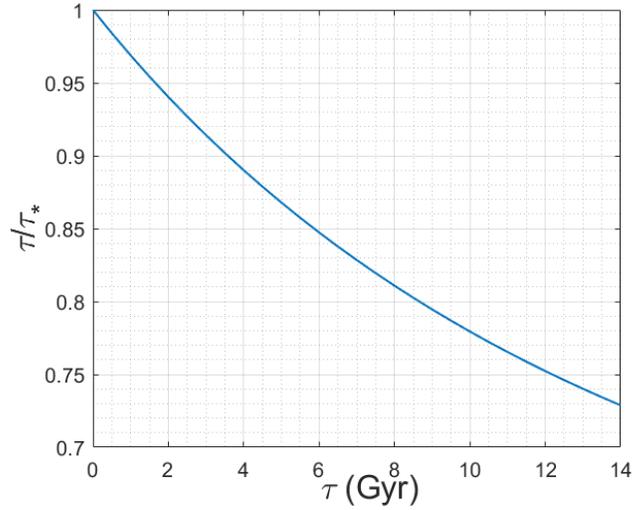

**Figure 2.** Variation of the ratio $\tau/\tau_*$ with stellar main-sequence age for the CCC model value of $\dot{G}_0/G_0 = 390 \times 10^{-12}$ yr$^{-1}$.

The findings of this study are summarized in Table 1. The $\dot{G}_0/G_0$ values correspond to the ratios $\mathcal{R} \equiv \tau/\tau_*$ for the models we have considered. The last row has only one ratio calculated for the $\dot{G}_0/G_0$ value predicted by the CCC model.

**Table 1.** The $\dot{G}_0/G_0$ values corresponding to the ratios $\mathcal{R} \equiv \tau/\tau_*$ for the models considered in this study: Degl'Innocenti's—linear variation model with $G(t) = G_0(1 + \Gamma_0(t-t_0))$, Degl'Innocenti's—powerlaw variation model with $G(t) = G_0(t/t_0)^\beta$, the CCC linear model with $c(t) = c_0(1 + c_1(t-t_0))$, and the CCC prediction $(c(t)/c_0)^{-1/2} = 1 + 0.0651(t_0 - t) - 0.0013(t_0 - t)^2$, as described in Section 3. The last row has only one ratio that has been calculated for the $\dot{G}_0/G_0$ value predicted by the CCC model.

| Model | $\mathcal{R} = \dfrac{\tau}{\tau_*}$ | $\dfrac{\dot{G}_0}{G_0} \times 10^{12}$ yr | $\mathcal{R} = \dfrac{\tau}{\tau_*}$ | $\dfrac{\dot{G}_0}{G_0} \times 10^{12}$ yr |
|---|---|---|---|---|
| Degl'Innocenti et al.—Linear |  | −35 |  | 7 |
| Degl'Innocenti et al.—Power-law | 0.5 | −6 | 1.5 | 6 |
| CCC—Linear |  | −429 |  | 429 |
| CCC—Prediction |  | Not Applicable | 0.77 | 390 |

## 4. Discussion

We have used the approach of Degl'Innocenti et al. [15] and modified it to be compliant with the CCC approach to explore if the variation of the gravitational constant predicted by the latter is consistent with the age of the globular cluster stars it determines. We applied the same assumptions for the *cosmological* and *local* energy conservation that we recently used to resolve the faint young Sun problem [61]. The interrelationship among the speed of light $c$, the gravitational constant $G$, the Planck constant $h$, and the Boltzmann constant $k$ of the CCC model, i.e., $G \sim c^3 \sim h^3 \sim k^{3/2}$, was used to relate the $c$ variation with the $G$ variation for comparing our findings with those of Degl'Innocenti et al. [15]. We determined that the CCC constraint on $\dot{G}_0/G_0$ is over fifty times laxer than that estimated by Degl'Innocenti et al. [15]. It is worth noticing that in addition to power law and linear variation of the constant consideres by Degl'Innocenti et al., we considered the variation of constants related through the function $\exp(a^{1.8} - 1)$.

The main reason for this discrepancy can be traced back to the application of *local* energy conservation laws to the *non-local* effects in general relativity and cosmology [61]. Energy conservation is not generally possible in general relativity and cosmology [68–74], which is also applicable to quantum mechanical systems [80]. Harrison [69,81] observed that the energy lost from one spatial region of a thermodynamically expanding Universe could not show up in another spatial region as work because, cosmologically, all regions are equivalent. Thus, all regions are expanding at the same rate. This led him to conclude energy is not conserved in the Universe. For a commoving ball of photon gas in the Universe, Peebles [68] considered energy loss inside it. He observed, "while energy conservation is a good local concept, …, and can be defined more generally in the special case of an isolated system in asymptotically flat space, there is not a general global energy conservation law in general relativity theory". An analysis of the deep conceptual problem encountered with the standard cosmological model in the context of the violation of energy conservation for commoving volumes was delineated by Baryshev [70]. His work also contains several vital references. Several non-conservation theories of gravity do not require null divergence of the stress–energy tensor [74]. The long and complex story relating to the debate on energy conservation in general relativity among some of the most brilliant minds—Felix Klein, Emmy Noether, David Hilbert, Albert Einstein, and others—was narrated by Brading [82].

In the work of Degl'Innocenti et al. [15] and other studies involving the evolution of stellar bodies to constrain the variation of gravitational constant e.g., [6,8,12,13,16,32], the potential energy change resulting from evolving $G$ contributes to the luminosity function. However, this energy change is *cosmological* and not available to affect the observable luminosity. Similarly, the energy change resulting from variations of $c$, $h$, or $k$ does not contribute to the observable luminosity. For example, even a meager 1% increase in the speed of light would cause the solar mass-energy to increase by 2%. If released as observable luminosity over the life of the Sun, the solar luminosity would increase by about an unacceptable two orders of magnitude over 5 billion years.

The energy non-conservation is evident when we allow energy density associated with the cosmological constant Λ to remain constant while the Universe expands. As a result, the total energy of the Universe increases without a concomitant decrease of any known source of energy. Similarly, one does not know where exactly the energy of a photon goes when it is redshifted in an expanding Universe. Thus, the energy conservation in cosmology must involve the sources and sinks of energy that are not directly observable, such as the vacuum energy of the quantum field theory. When one ignores this requirement, it naturally leads to unrealistic constraints on the variation of the constants.

Just as a side note in the context of globular clusters, it is worth mentioning briefly the work of Dearborn and Schramm [14]. They estimated the effect of $G$ variation on the dimensions of these clusters for constraining $G$. They considered in their study how the self-binding energy of a cluster would evolve when $G$ is not constant and consequently affect the dimensions of a cluster. Once the variation of $c$ and the effect of $c$ variation on the unit of length are considered, the kinetic energy is affected precisely with the same proportionality as the potential energy. The net result is that the dimensions of the clusters are not affected by the variation of the constants, and this fact cannot be used for constraining $\dot{G}/G$.

The challenge remains to come up with a controlled experiment, astrometric or terrestrial, that could unequivocally prove the variation of any of the four constants without constraining others to their fixed currently known values. Since the Planck constant is already possible to measure with a precision of about 10 parts in a billion using the Kibble balance e.g., [83–86], it is obvious to consider if it can be improved so that when measured over say a one-to-ten-year period, it could constrain the variation of $h$. We have explored this possibility [87] as $h$ is used now to define the unit of mass and weighing of a mass in the Kibble balance involves $c$, $G$, and $h$ through mass $m$ which is proportional to $h/(Gc^2)$ [88]. XRCD (x-ray crystal density) method can also determine the Planck constant [89]) with precision similar to the Kibble balance. Another high precision method of measuring $h$ is based on photoemission spectroscopy [90,91]. The precision of this method is currently lower than the Kibble balance and XRCD method but may have the potential to increase significantly by improving the design of the electron spectrometer and other components of the apparatus. Another possibility is a high precision determination of the Boltzmann constant [92,93] which can already be measured with an accuracy approaching 1 part in a million. In these experimental methods and any others that may be conceived, it will need to be ensured that one takes into account the variation of other constants that might directly or indirectly influence the outcome of the measurements.

**5. Conclusions**

The ages of the main-sequence stars in globular clusters are not in conflict with those determined using the co-varying coupling constant (CCC) approach, provided:

1. Energy conservation includes the *cosmological* energy along with the *local* energy.
2. Variations of gravitational constant $G$, the speed of light $c$, the Planck constant $h$, and the Boltzmann constant $k$ are considered interrelated as $G \sim c^3 \sim h^3 \sim k^{3/2}$.

The variations of $G$ and $c$, i.e., $\dot{G}/G = 3\dot{c}/c = 390(\pm 4) \times 10^{-10}$ yr$^{-1}$, determined in earlier publications are consistent with the present study. However, the challenge is to design an experiment, astrometric or terrestrial, which could, without doubt, establish the variation of any of the four constants without fixing remaining to their known laboratory-measured values.

**Funding:** this research was partially funded by Macronix Research Corporation research grant2020-23.

**Institutional Review Board Statement:** Not applicable

**Informed Consent Statement:** Not applicable

**Data Availability Statement:** No external data was used in this research.

**Acknowledgments:** The author is grateful to Barry Wood of NRC (National Research Council of Canada) for an informed discussion on the limitation of the Kibble balance for measuring the Planck constant at a precision better than ten parts in a billion. His thanks are also due to Stephan Schlamminger of NIST (National Institute of Standards and Technology, USA) for considering the possibility of testing the variability of constants with the most advanced NIST Kibble balance. He wishes to express his most sincere gratitude to the reviewers of the paper whose constructive critical comments were instrumental in greatly improving the quality and clarity of the paper.

**Conflicts of Interest:** The author declares no conflict of interest.